# Photoacoustic Image Quality Improvement from a Single Cell Low Frequency PMUT

Kaustav Roy[a,c#], Arijit Paramanik[b#], Souradip Paul[b], Akshay Kalyan[c], Eshani Sarkar[c], Anuj Ashok[d], Rudra Pratap[c*], M Suheshkumar Singh[b*]

[a]Electrical and Computer Engineering, National University of Singapore, Singapore 117583,
[b]School of Physics, Indian Institute of Science Education and Research, Thiruvananthapuram 695551, KL, India,
[c]Centre for Nano Science and Engineering, Indian Institute of Science, Bangalore 560012, KA, India,
[d]Elmore Family School of Electrical and Computer Engineering, Purdue University, West Lafayette 47907, IN, USA

\* All correspondence should be addressed to suhesh.kumar@iisertvm.ac.in and pratap@iisc.ac.in
\# Authors made equal contribution

## ABSTRACT

Photoacoustic image (PAI) quality improvement using a low frequency piezoelectric micromachined ultrasound transducer (PMUT) having the fundamental resonant frequency ~ 1 MHz is being reported. Specifically, three different methods are implemented such as – the frame averaging, mathematically improved algorithms, and a hardware position accurate arrangement in order to obtain unparallel PAI image quality. Validation study has been conducted in both agar phantom and ex-vivo tissue samples. Measurable image quantifiers in the form of full width at half maximum (FWHM), signal to noise ratio (SNR), and contrast ratio (CR) are used to evaluate the improvement in the image quality. It is found that the FWHM increases by 34%, SNR by 23% and CR by 25%, suggesting the efficacy of the methods to achieve better photoacoustic images employing PMUT-based detector. The study demonstrates that the suggested methods of improvement could play a key role in a promising cost-effective PMUT-PAI system in future.

**Keywords:** PMUT, Photoacoustic Imaging, nanofabrication, thin film, MEMS

PMUTs are a class of powerful based micro acoustic devices which are driven by thin film piezoelectricity [1]. These devices are special since they can be made tiny, in any form factor, have minimal power consumption, can be made CMOS compatible, and can be manufactured in bulk batches [2]–[12]. They are fabricated by using precise nanofabrication [13], [14] tools which gives them a sheer benefit over the ubiquitous bulk ceramic-based ultrasound transducers, since the accuracy of such tools to create the tiniest design implementation is unparallel. There has been a recent trend in using such devices for functional imaging of deep tissue by using the photoacoustic (PA) modality [15], [16] – which is fundamentally based on boundary detection of short-duration light induced ultrasound signal by an ultrasound receiver. This can be attributed to the above-mentioned advantages and there have been several contributions in recent years towards this direction [10], [17]–[23]. While there have been considerable efforts in generating PA images using high frequency PMUTs (operational fundamental resonant frequency > 5 MHz), with several of the cells connected together to form an element, images from a low frequency (~ 1 MHz) PMUT having a single cell element is yet unreported till date. This is due to the inability of the low frequency ultrasound receivers to resolve finer details [24] owing to their longer wavelengths. In this contribution we aim at solving such a limitation by using a combination of backend signal processing, reconstruction algorithms and hardware positioning to resolve targets with an improved exactness, thereby breaking the wavelength-resolution limit. This image quality improvement has two important implications – (a) high quality image can be obtained from deeper depths with relative ease in device microfabrication, since fabricating low frequency PMUTs by bulk silicon micromachining is easier as compared to high frequency PMUTs due to a shallower etch aspect ratio; (b) the fact that a single cell PMUT works as the PA receiver can be leveraged to create high density arrays with a much-improved form factor, which will in-turn improve the tomographic image quality resulting from such arrays.

Firstly a photoacoustic module was created by fabricating a 3D printed preform containing a provision for two optical posts, a snug fit slot for the printed circuit board (PCB) mounted PMUT and the optical output from a laser as shown in Figure 1 (a(i)). The PMUTs used for this work were fabricated at the National Nanofabrication Centre, Centre for Nano Science and Engineering, Indian Institute of Science, Bangalore, on a Silicon-on-Insulator (SOI) substrate using thin film (thickness ~ 800 nm) lead zirconate titanate (PZT) as the active layer. A standard fabrication recipe is followed [25] in order to realize the devices. Each PMUT chip contains a single cell device (see Figure 1 (a(ii))) having a stack thickness of 10 μm, and has a circular geometry, having a diameter of 250 μm. A scanning electron micrograph (SEM) captures the finer details of the single cell PMUT as shown in Figure 1 (a(iii)). A PA imaging phantom was subsequently prepared which contained a single embedded graphite target of circular geometry (diameter: 500 μm) as shown in Figure 1 (b). Subsequently, a pulse photoacoustic instrumentation setup was constructed as shown in Figure 1 (c) by assembling a nanosecond pulsed laser (SpitLight EVO S OPO, Innolas Lasers, Germany) of wavelength 532 nm, pulse width ~ 6 ns, and pulse repetition frequency of 100 Hz, the output of which was connected to a laser fiber cable to direct the light into the photoacoustic target. The phantom was thus placed in a suitable position to achieve the maximum illumination condition. The photoacoustic module was mounted on a 3-axis motorized stage, which was connected to a computer for precise control of the light and the PMUT's position. The phantom and the PMUT were submerged under water, contained in a fish tank for efficient acoustic coupling. The laser pulse was allowed to illuminate the target (having fluence ~ 10 mJ/cm$^2$), which then generated photoacoustic (PA) waves from the target and were received from the PMUT in the form of time variant electrical charge. The PMUT's small signal is then amplified by using a pulser-receiver (5073PR-40-P, Olympus) and sample acquired using a data acquisition card (779745-02, NI PCI-5114, National Instruments, 250MS/sec) attached to the computer. The data is stored on the computer for further processing. Next, preliminary characterizations were carried out to understand the performance of the PMUT and the PA module. The PMUT was thus wire-bonded to the custom-made PCB which was terminated in low noise SMA connector to establish an efficient electrical connection. The fundamental frequency of the device is found by using the laser doppler vibrometer (MSA 500, Polytec Inc.) to be 1.63 MHz in air with a maximum displacement of 532 nm when excited with 1 V AC as shown in Figure 1 (d(i))). Next, in order to understand the electrical characteristics of the device, the impedance analyzer was employed which depicted the device impedance to be ~ 1.97 kΩ at 1.63 MHz, with a phase shift of ~ 6° at resonance (see Figure 1 (d(ii))). Next, the PMUT was characterized for the PA A-line under water and the received signal was plotted and is depicted in Figure 1 (d(iii)). In water the PMUT exhibits significant virtual mass loading [3], and the resonant frequency gets lowered to ~ 1.05 MHz. The plot is further analyzed to calculate the -6 dB fractional bandwidth which was found to be 70 %. Now, in order to form a PA image from a target, five different steps are followed as shown in Figure 1(e). First, the PMUT is scanned over a designated space and raw PA signals are acquired. Next, a beamformer is applied on the raw signals followed by the envelope detection which is hence fed to log compression to finally display the PA image.

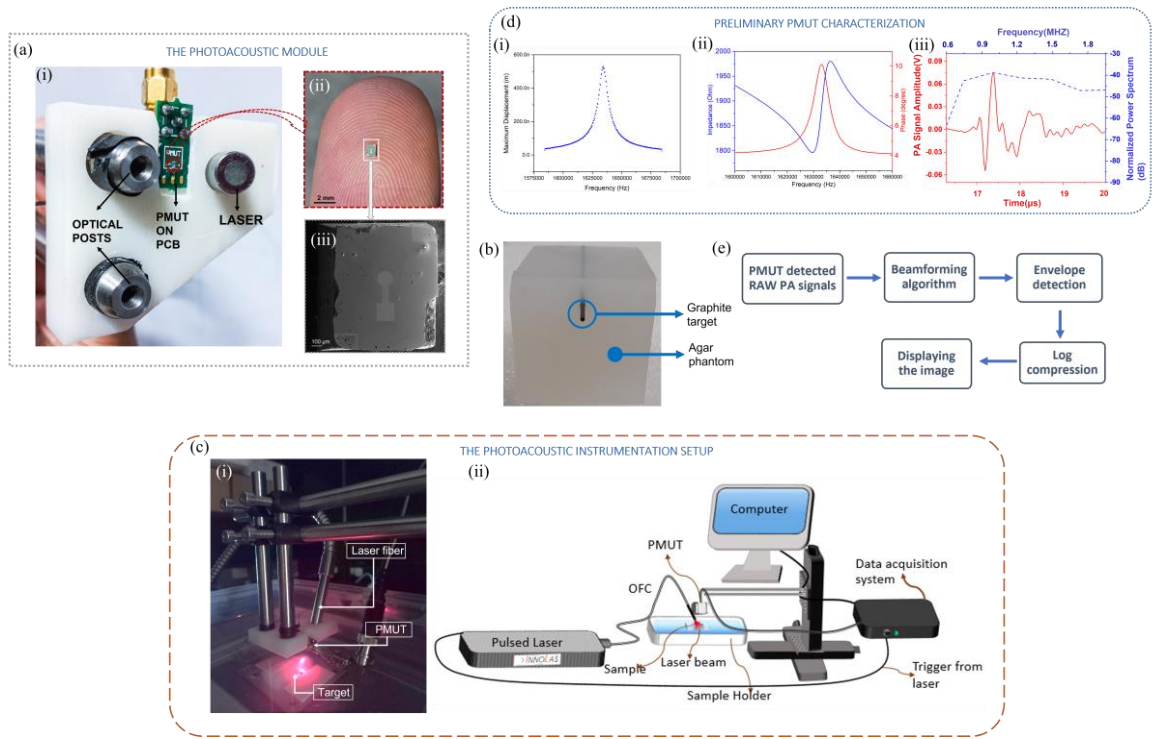

Figure 1. The photoacoustic system used in the work: (a) (i) The photoacoustic module consisting of a 3D printed preform onto which the PMUT mounted on a PCB and the laser fiber head is attached to at an angle, (ii) The PMUT used in the work held on a finger, (iii) Magnified scanning electron micrograph of the PMUT (b) Agar phantom prepared with a single graphite target embedded inside (c) The photoacoustic instrumentation setup, (i) Snapshot while laser pulsing, showing the fiber head, PMUT on the preform and the target, (ii) Schematic of the experimental setup used for the work (d) Preliminary PMUT characterization, (i) laser doppler vibrometer of the PMUT showing a resonant frequency of 1.64 MHz in air, with a maximum displacement of 510 nm at 1 V, (ii) electrical impedance measurement showing an impedance of 1900 Ω at the resonant frequency with a phase shift of 6 degrees, (iii) Photoacoustic A-line time and frequency data as obtained from the agar phantom having a -6 dB fractional bandwidth of 70 % (e) Flow chart representing steps involved to form a B-mode PA image from the data matrix.

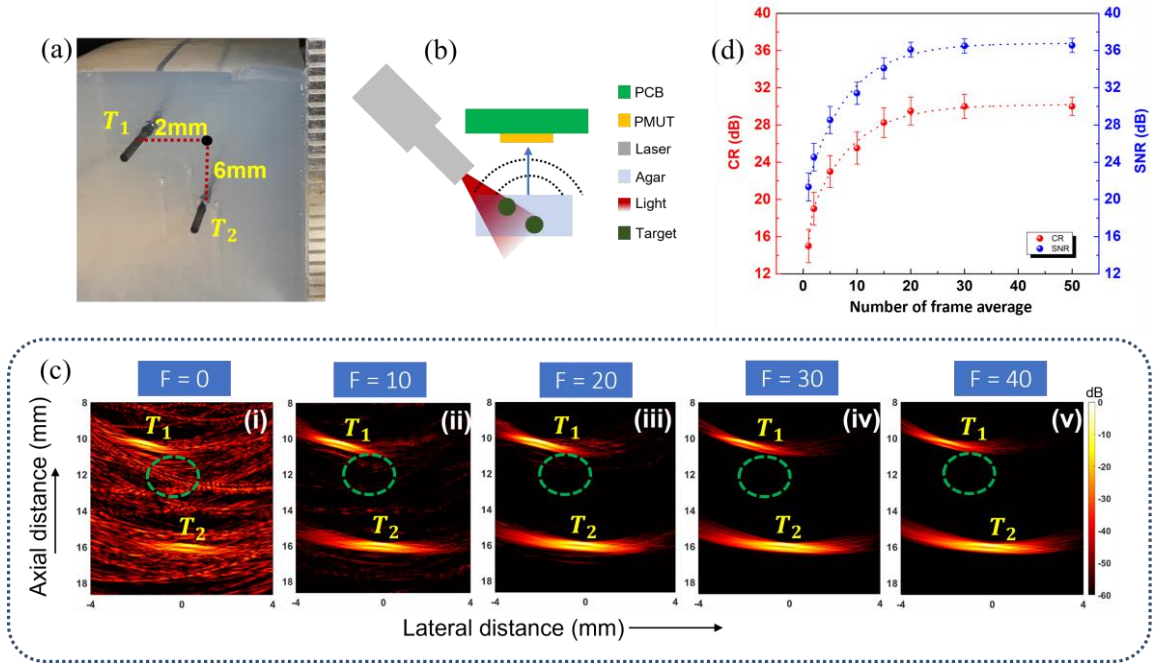

Figure 2. Effect of frame average on the quality of PA image: (a) Agar phantom prepared with two graphite targets embedded inside (b) Schematic of the photoacoustic arrangement (c) PA images obtained from different frame averages from 0 to 40. (d) CR and SNR plotted with respect to the number of frame average both of which is observed to get saturated after 20 averages.

Next, an agar-based phantom containing two graphite targets in a diagonal relationship Figure 2 (a) is prepared and illuminated with the nanosecond laser pulse. The PA module kept in a specific fashion Figure 2 (b) received the ultrasound to produce A-line data from the associated section of the phantom. The PMUT is then made to scan a definite area of interest at a step size of 50 μm, thereby generating an array of A-line data. The array is defined as a frame, which when passed through the delay and sum (DAS) reconstruction algorithm to form an image. The effect of averaging multiple frames from the area of interest is then studied, and it is found that frame averaging has a profound effect on the image quality so obtained Figure 2 (c). To quantitatively measure the improvement in the quality of image, the signal to noise ratio (SNR) and the contrast ratio (CR) is calculated from each image by comparing the average intensity pixel values at the target region and the background region (circle in green) [26] and is reported in Figure 2 (d). It is observed that the SNR and CR follow a similar trend and improve by 15.5 dB and 15 dB respectively. It is also observed that both the SNR and CR saturate after an average of 20 frames which thus can be considered the optimized number effective enough to drive off the combined effects of random ambient electromagnetic noise, noise due to on-chip trace and bond parasitic capacitance and additional noise from the receive signal amplifier.

Next, in order to investigate the performance of PMUT based PA imaging, various beamforming algorithms [27] were evaluated on PMUT detected PA signals from the agar phantom containing four graphite targets as shown in Figure 3 (a). The PA module kept in a specific fashion Figure 3 (b) received the ultrasound to produce A-line data from the associated section of the phantom. Delay and Sum (DAS) is one of the most common beamforming techniques due to the simplicity involved in implementation. Figure 3 (c(i)) represents the reconstructed targets using the DAS algorithm. However from the image it is observable that the targets are corrupted with side-lobes. This is due to the non-adaptiveness and blindness nature of DAS beamformer. To overcome this limitation, Coherence Factor (CF) based adaptive weighting technique is used. DAS combined with CF provides a significant improvement in the reconstructed images as shown in Figure 3 (c(ii)). As can be observed, the side lobes and the background noise is reduced to a great extent. Targets are clearly detectable at all the imaging depths. This is due to enhancement of coherent contribution of the beamformed signals [28]. Next, a non-linear Delay Multiply and Sum (DMAS) beamformer algorithm was implemented in the same data set. It is observed that DMAS shows better improvement (see Figure 3 (c(iii))) as compared to the native DAS beamformer in the same data set. Combinatoric multiplication of delayed signal intensifies the coherent signal contribution. Simultaneously, the noises arise from incoherent parts and sidelobes are diminished, as a result image quality is remarkably improved. Moreover, by using DMAS, it is possible to better capture the target aspect ratio as compared to the DAS + CF beamformer. Finally, a modified version of CF (CF') [29] was implemented together with the DMAS to get the image as shown in Figure 3 (c(iv)), and is observed to produce the best PAI in terms of the reduction in side lobes and preservation of the target aspect ratio due to the efficacy of DMAS in the numerator of CF instead of traditional DAS [27]. Once the efficacies of various beamformers are assessed qualitatively, a quantitative analysis of the PA images is undergone as shown in Figure 3 (d), (e). First, the normalized pixel amplitude is plotted with respect to the lateral distance (see Fig 3(d)) for the target 't2' at an imaging depth of 1.15 cm for the above mentioned beamformers. It was found that the side lobes reduce from -4 dB to -16 dB and the lateral spread at -3 dB pixel intensity reduces from 1.58 mm to 0.59 mm for DAS beamformer to DMAS + CF' beamformer respectively. In order to further analyze the PA images, quantitative bar plots were obtained in terms of three different measurables with respect to the imaging depth – Full Width at Half Maximum (FWHM), SNR and CR. Figure 3 (e(i)) depicts that FWHM is strongly dependent on the kind of beamformer used. With depth FWHM was found to vary by 34 % for DAS and 30 % for DMS + CF', suggesting that the later beamformer is more efficient in terms of producing lesser variation in the FWHM with respect to depth. Figure 3 (e(i),(ii)) depicts the impact of various beamformers on the targets' SNR at various depths. A general trend of nonlinear increase in the SNR and CR is observed with an increase in the imaging depth. A comparison for the various beamformers depicts that DMAS + CF' has the best SNR and CR of 39 dB as compared to all other beamformers.

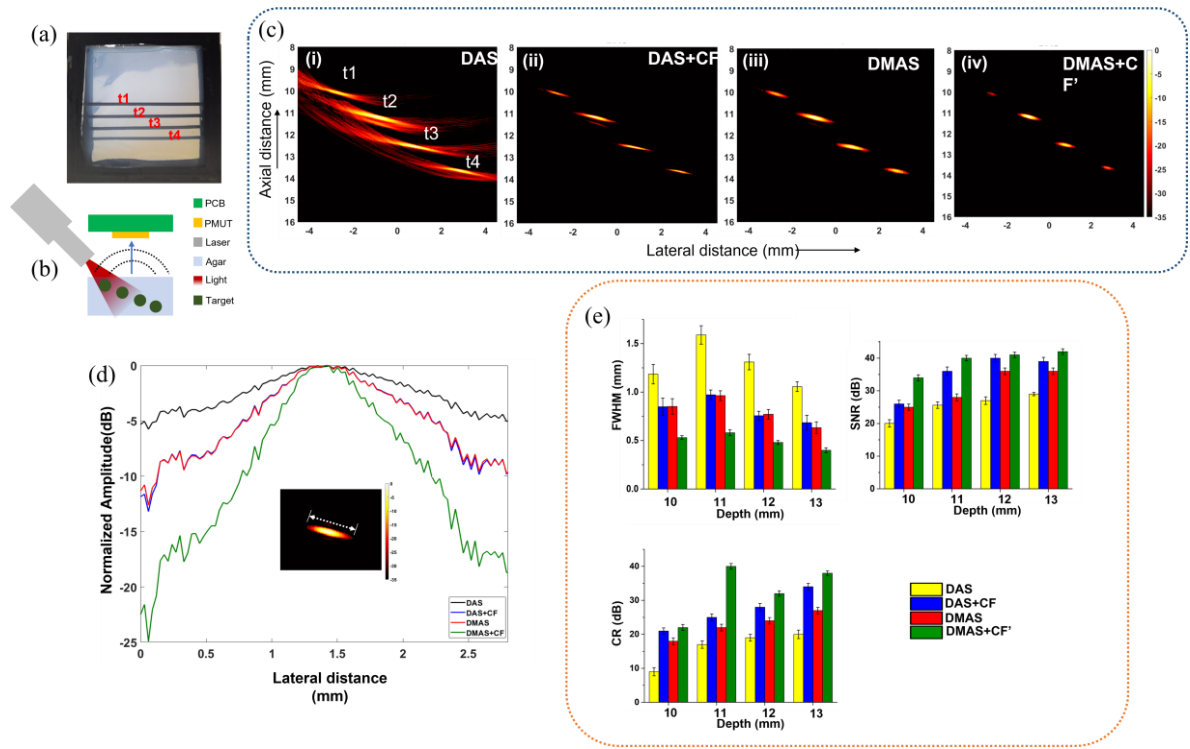

Figure 3. Effect of various advanced beamformers on the quality of PA image: (a) Agar phantom prepared with two graphite targets embedded inside (b) Schematic of the photoacoustic arrangement (c) PA images obtained from various algorithms (i) DAS (ii) DAS + CF (iii) DMAS and (iv) DMAS +CF' (d) Normalized amplitude plotted with respect to the lateral distance (e) Quantitative plots in form of FWHM, SNR, and CR with respect to depth.

In the next study, the effect of the hardware positioning system as shown in Figure 4 (a), specifically the alignment of the target with respect to the axes of the laser illumination and the PMUT on the quality of PA image formed is studied [30]. First an agar phantom is prepared containing a hair (diameter ~ 200 μm) which served as the PA point target (see Fig 4 (b)). As shown in Figure 4 (a), the tilt of the laser light with respect to the horizon (θ) and the center-to-center distance between the light and the PMUT (d) is held constant. This is imperative while performing real-time photoacoustic tomography since the design of the light integrated ultrasound transducers is limited by hard constraints in terms of space. The correct position of the PMUT and light ray is thereby calculated and either of their positions thus changed as depicted in Figure 4 (c). The gap between the PMUT and the target is enhanced from 14 mm to 18 mm in the interval of 1 mm and the position of the laser changed accordingly in order to coincide the respective axes with the target. An improvement in the image quality is observed in terms of the tilt of the target and its spread. The FWHM from each image is calculated and reported in Figure 4 (d). It is observed that the FWHM decreases from 0.8 mm to 0.35 mm. This observation can be reasoned as follows – as shown in the Figure 4 (a), when the laser illuminates the target in its axis and the PMUT is out of the on-axis reception, there is an additional time delay involved for the sound to propagate from the target to the PMUT. This extra delay is not accounted for in the beamforming algorithm which elongates and tilts the target's image. When the target is in the exact intersection position of the illumination and the PMUT's axes, the additional delay is negligible which results in the accurate beamforming implementation, thereby improving the quality of the PA image formed.

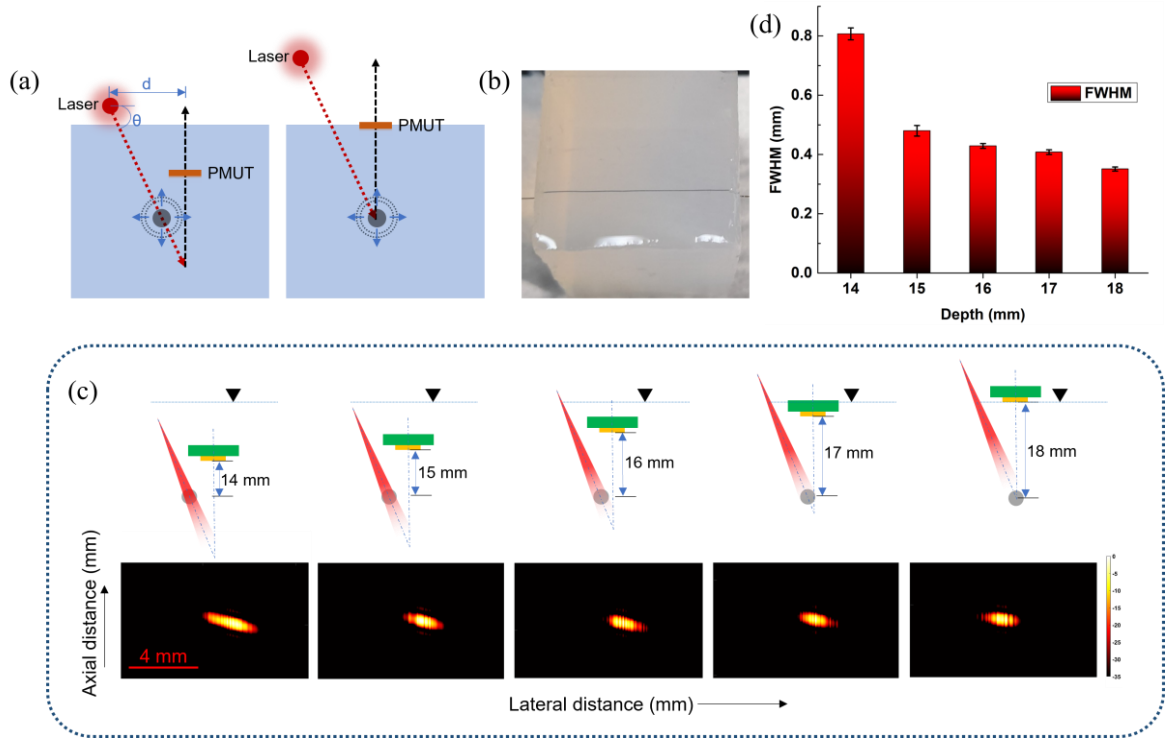

Figure 4. Effect of hardware positioning system on the PA image quality: (a) Graphical representation of the proposed hardware positioning with two cases – (i) Laser colinear with the target, but PMUT out of axis, (ii) Laser and PMUT placed such that both of them are colinear with the target (b) Agar phantom prepared with a single strand of human hair embedded inside (c) Effect of change in the position of the laser and the PMUT on the quality of PA image (d) Dependence of FWHM on the position of the photoacoustic module with respect to the target

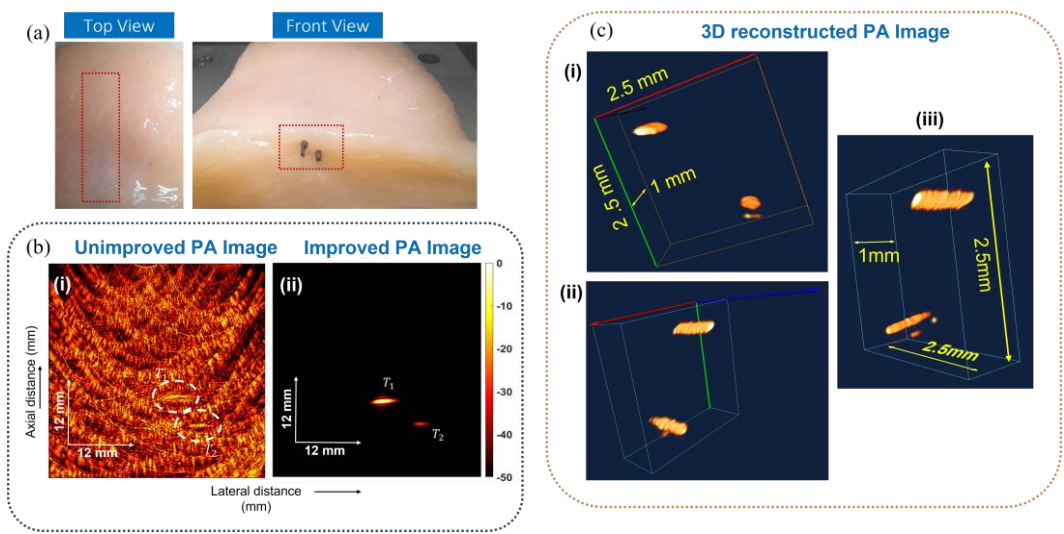

Figure 5. Effect of combination of the three methods on the PA image quality: (a) Chicken breast phantom prepared with two graphite targets embedded inside (b) PA Image formed from two contrasting situations (i) unimproved image obtained from the data matrix after applying DAS beamformer, (ii) improved image obtained from the data matrix after applying 20 frame averages + (DMAS + CF') + accurate hardware positioning (c) 3D reconstructed PA image obtained from the phantom

In order to validate the effect of the combination of all the proposed methods on the PA image from a realistic specimen, experiments were conducted on chicken breast tissue (length x width x thickness: 30 mm x 30 mm x 15 mm) which was loaded with two graphite targets of 500 μm, placed diagonally as shown in the Figure 5 (a). The A-line signals picked up by raster scanning the phantom with the photoacoustic module was (a) beamformed with the DAS beamformer as a control (see Figure 5 (b(i))) in contrast to (b) frame averaged for 20 times, beamformed with the DMAS + CF' beamformer, and hardware position corrected (see Figure 5 (b(ii))) in order to observe the improvement in the PA image quality. Thereafter, a 3D PA image was also captured by scanning the phantom by performing a 2D raster scan of the phantom, and the image formed is being reported in Figure 5 (c). The graphite targets are clearly visible with negligible background noise.

In conclusion, this contribution targeted at obtaining good quality PA images from various targets by using a home built pulsed laser-based PA system in conjunction with a single cell low frequency PMUT. The effects of a combination of three methods such as – averaging of imaging frames, advanced beamformers, and accurate hardware positioning was experimented with, and the results are demonstrated herein. It is perceived that the combination of such methods has profound impact on the PA image so formed and can greatly improve the quality to several orders in terms of improvement of the correct dimension realization in terms of FWHM, signal to noise and contrast to noise ratios, and unnecessary tilt associated with imaging targets. Briefly, the presented methods could be a prime factor that can extend the horizon of PMUT-based PAI in terms of image quality improvements. Above all the proposed method is easily implementable without employing any advanced hardware platform.